\documentclass{icrc29}
\usepackage{graphicx,amssymb,amsmath,times}
\begin{document}
\title [Factorial moment ...]{Factorial moment studies of Cherenkov images}
\author[A.Razdan] {A.Razdan, D.K.Koul, M.K.Koul, S.K.Charagi\\
           Nuclear Research Laboratory,Bhaba Atomic Research Centre,
            Mumbai 400 005, India \\
        }
\presenter{Presenter: A. Razdan (akrazdan@apsara.barc.ernet.in), \
ind-razdan-A-abs1-og27-poster}

\maketitle

\begin{abstract}
In this paper we study factorial moments of simulated  Cherenkov images of
TACTIC type $\gamma$-ray telescope for cosmic $\gamma$-ray  and proton segregation.
A comparative  and supplementary studies between factorial moments  and
Hillas parameters using ANN is discussed.
\end{abstract}

\section{Motivation}

In this exploratory simulation studies, we use
factorial moments as a tool to  segregrate  $\gamma$-ray 
and proton initiated showers.
Extensive Air Showers (EAS) are produced when VHE/UHE primary
photons, protons and other high Z nuclei enter atmosphere from the
top and produce Cherenkov radiation which can be studied by imaging
technique or by measuring lateral distribution. 
In the Hillas parameterization approach, image is approximated as 
an ellipse and parmeters like image  "shape" and "orientation" are calculated. 
Hillas paramaters like $\it length$,$\it width$,$\it alpha$,$\it distance$,$\it size$ etc.
are essentially set of second order moments. Among these parameters
"width" and "alpha" are known to  be very powerful parameters for segregation of
$\gamma$-ray initiated cherenkov photon from cosmic ray  background.
Using this approach, present day $\gamma$-ray telescope have been able
to reject cosmic ray background events up to 99.5 $\%$ level,
while retaining upto 50 $\%$ of the $\gamma$-ray events from a point
source. Hillas parameters are found to be good classifiers for
small images (close to telescope threshold energies) but
fail for large images (of higher primary energy) as too many
pixels are part of the image. Again for GeV energy region telescopes like MAGIC and
MACE, Hillas parameters alone  are not very attractive. A loss of 50 $\%$ of actual
image due to parameterization is also a big constraint. 
Hence the idea of trying factorial moments as a possible tool
is a subject matter of this paper.

\section{ Global factorial moments}

A given image is divided into M equal bins of width d. If $n_m$ is the
number of particles in the mth bin, quantity $s_i$ can be defined as
\begin{equation}
(s_i)_m = n_m (n_m -1)........(n_m -i+1)
\end{equation}
and calculated for each bin.
The standard moments are defined as
\begin{equation}
F_i ^s(M)=\frac{< \frac{1}{M} \sum_{m+1} ^{M} (s_i)_m> M ^{i}}{< N >^{i}}
\end{equation} 
The normalized factorial moments are 
\begin{equation}
F_i ^n (M)=\frac{ \frac{1}{M} \sum_{m+1} ^{M} (s_i)_m M ^{i}}{ N(N-1).....(N-i+1)}
\end{equation}
where N is the total number of photoelectrons in the image.
Factorial moment has a power law dependence 
\begin{equation}
< F_i ^n (M) > \sim ( \frac{1}{d}) ^{\alpha}
\end{equation}

\section{ Simulation strategy}

For the present studies simulatons were carried out using CORSIKA [1](version 5.6211) along  with
EGS4, VENUS,GHEISHA codes for Chenenkov option. Simulated data is generated for 208 elements
of TACTIC [2] configuration, each element of the size 4m $\times$ 4m. Simulated data corresponds to
Mt.Abu altitude ( 1300 m)  and appropriate  magnetic field. Cherenkov databases correspond to
wavelength band of 300-450 nm. Simulated databases have been subjected to atmospheric absorption.
For realistic simulation studies of TACTIC array, photoelectron content of each pixel is modified
to include shot noise contribution from night sky background. A noise profile based on actual experimental
conditions is superimposed on each image.

\begin{figure}{h}
\begin{center}
\includegraphics*[width=0.8\textwidth,angle=0,clip]{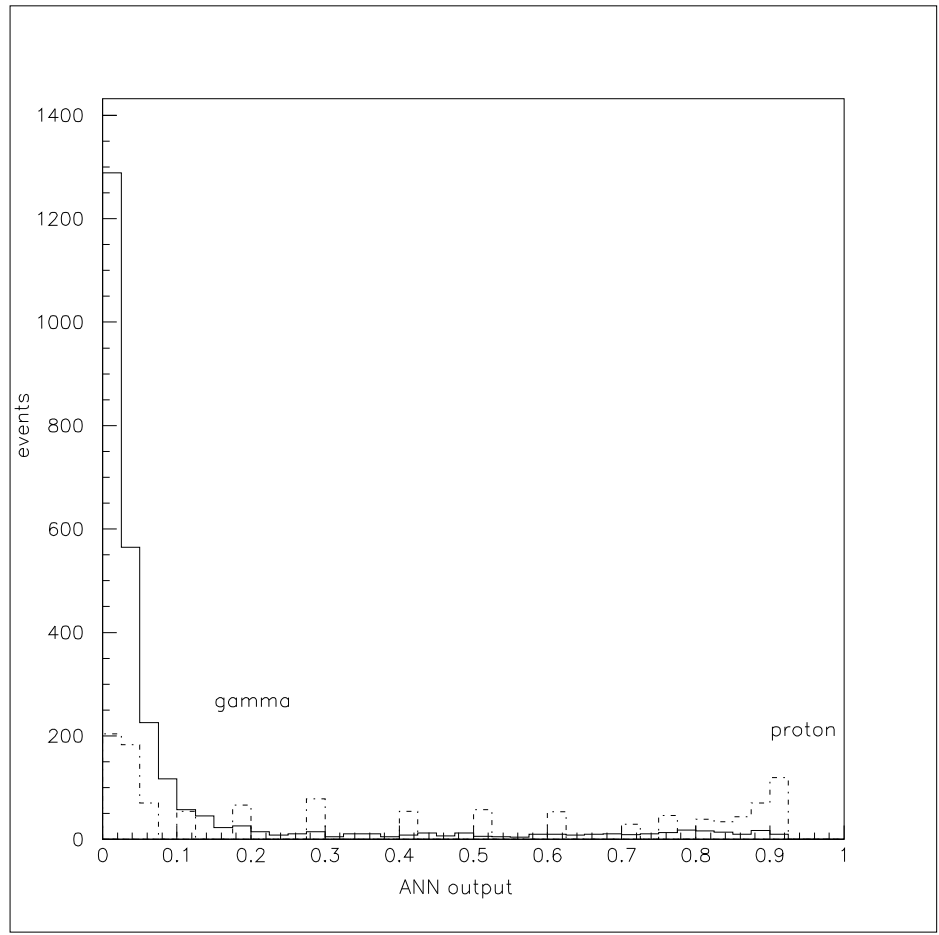}
\caption{ANN output with Hillas parameters as input}
\end{center}
\end{figure}

In this studies 15800 cherenkov 
images each corresponding to $\gamma$-ray (1 TeV)  and proton (2 TeV) initiated showers
are considered. The innermost pixels of the camera generate low level 
gamma ray threshold triggers using 3NCT ( Nearest neighbour non-collinear triplets)
topological trigger configuration [3]. Only those 
images are selected  which satisfy 3NCT trigger configuration with the condition that there are minimum of 6
photoelectrons in each pixel. By applying 3NCT criteria a total of 5300 $\gamma$-ray images and
2407 proton images are obtained. Hillas parameters and factorial moments have been obtained for these chosen
images. 

Cherenkov images of $\gamma$-ray showers are mainly elliptical in shape, hence compact. However, the cherenkov
images of hadronic showers are mostly irregular  in shape.  
For all cherenkov images, shape parametes like length, width and distance and orientation parameters like 
$\alpha$, azwidth and Miss are calculated.  It is observed that width and $\alpha$ parameters 
have strong segregation capability.For all cherenkov images 
factorial moments $F_i^n$ of order i (i=2,3,4,5,6) have been calculated.. It is observed that factorial
moment of fifth and sixth order are very powerful parameters to segregate $\gamma$-rays from protons.
We have used Artificial neural network (ANN) for event classification. Figure 1 shows ANN output with only
Hillas parameters (width and $\alpha$) as ANN inputs. Figure 2 displays ANN output with  fifth 
and sixth order factorial moments as ANN inputs.
Only those images with 
fifth order and sixth order moments greater than 4000 have been considered. Figure 3 depicts ANN output with
hillas parameters and factorial moments (5th and 6th order only) together as ANN inputs.

\begin{figure}[h]
\includegraphics*[width=0.8\textwidth,angle=0,clip]{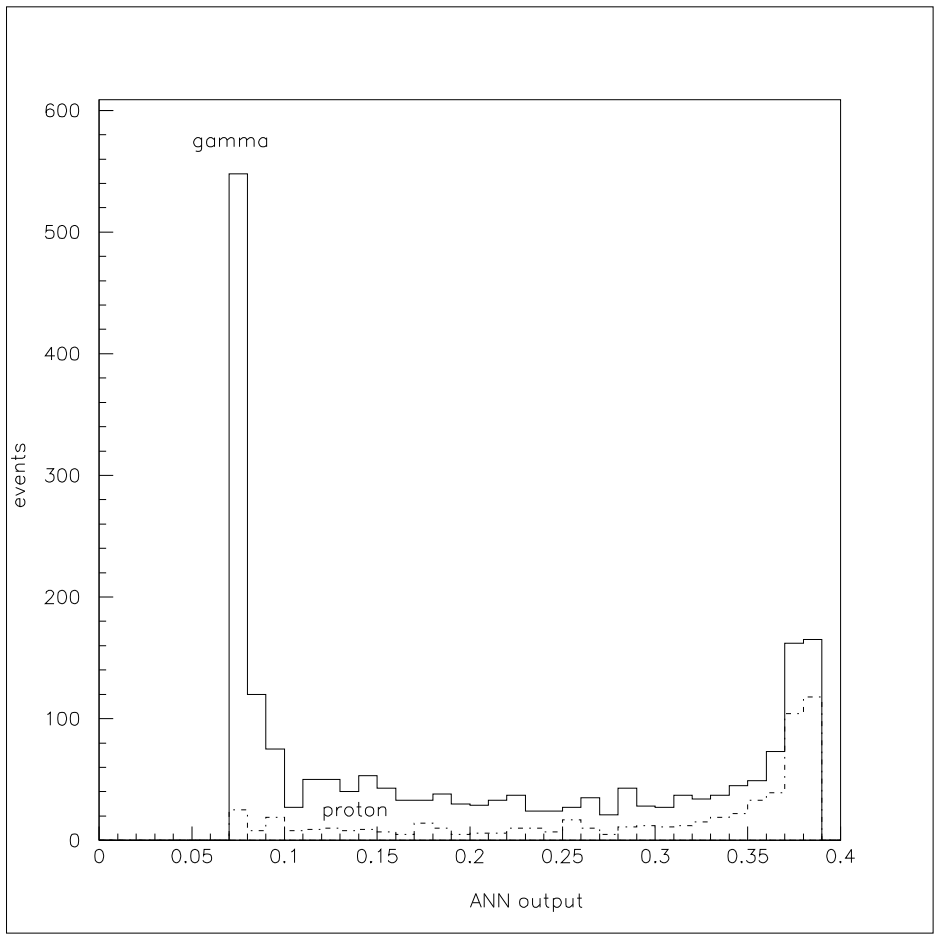}
\caption{ANN output with factorial moments as input}
\end{figure}

\section{conclusion}
It is clear that factorial moments are very powerful parameters and combine very well with hillas
parameterization  for segregation of $\gamma$-rays from protons.
\begin{figure}[h]
\includegraphics*[width=0.8\textwidth,angle=0,clip]{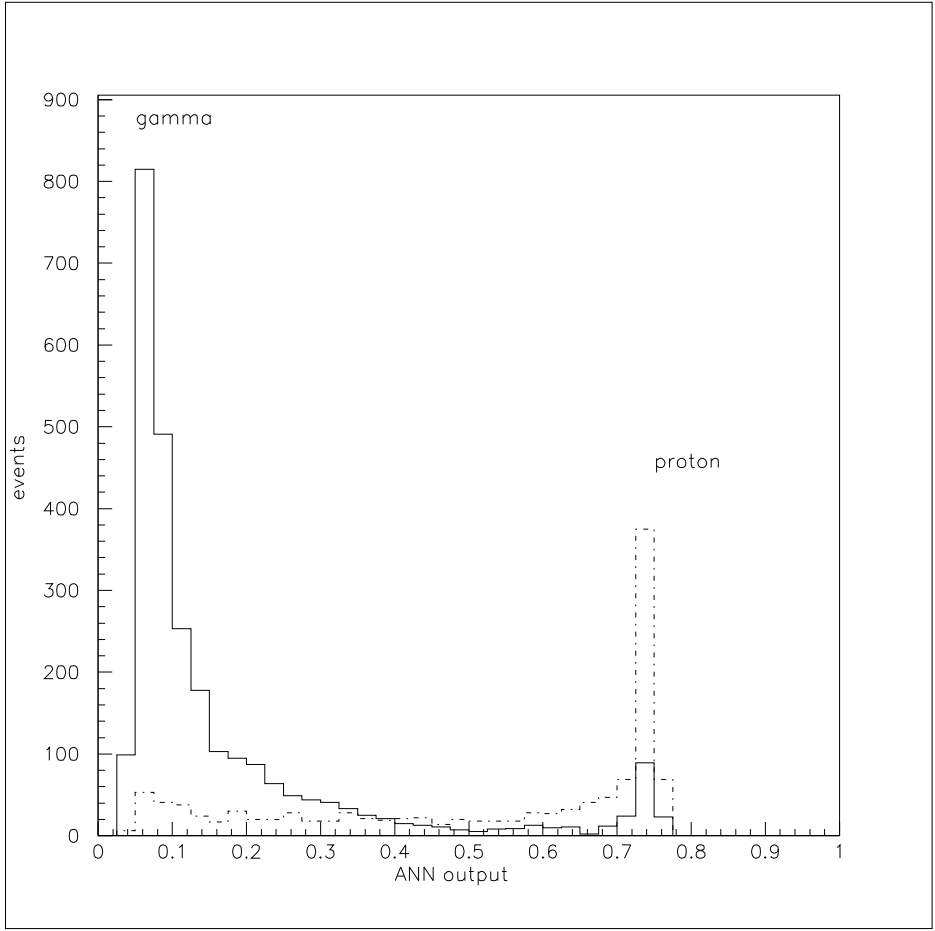}
\caption{ANN output with Hillas paramets and factorial moments as input}
\end{figure}

\end{document}